\documentclass{article}

\usepackage{arxiv}

\usepackage[utf8]{inputenc} 
\usepackage[T1]{fontenc}    
\usepackage{hyperref}       
\usepackage{url}            
\usepackage{booktabs}       
\usepackage{amsfonts}       
\usepackage{nicefrac}       
\usepackage{microtype}      
\usepackage{lipsum}         
\usepackage{graphicx}
\usepackage[numbers]{natbib}
\usepackage{doi}
\usepackage[nolist]{acronym}
\usepackage{amsmath}
\usepackage{amsfonts}
\usepackage{amssymb}
\usepackage{mathbbol} 

\DeclareSymbolFontAlphabet{\mathbb}{AMSb}
\DeclareSymbolFontAlphabet{\mathbbl}{bbold}

\newcommand{\eg}{\emph{{e.g., }}}
\newcommand{\ie}{\emph{i.e., }}
\newcommand{\figref}[1]{Fig.~\ref{#1}}
\newcommand{\tabref}[1]{Table~\ref{#1}}

\title{Optimum Network Slicing for Ultra-reliable Low Latency Communication (URLLC) Services in Campus Networks}
\date{December 23, 2022}

\newif\ifuniqueAffiliation
\uniqueAffiliationtrue

\ifuniqueAffiliation 
\author{{Iulisloi Zacarias, Francisco Carpio, André Costa Drummond, Admela Jukan} \\
	Institut für Datentechnik und Kommunikationsnetze\\
	Technische Universit\"at Braunschweig\\
	Braunschweig, Germany \\
	\texttt{\{i.zacarias, f.carpio, andre.drummond, a.jukan\}@tu-braunschweig.de} \\
}
\else
\usepackage{authblk}

\newbox{\orcid}\sbox{\orcid}{\includegraphics[scale=0.06]{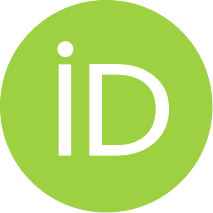}} 
\author[1]{%
	\href{https://orcid.org/0000-0000-0000-0000}{\usebox{\orcid}\hspace{1mm}David S.~Hippocampus\thanks{\texttt{hippo@cs.cranberry-lemon.edu}}}%
}
\author[1,2]{%
	\href{https://orcid.org/0000-0000-0000-0000}{\usebox{\orcid}\hspace{1mm}Elias D.~Striatum\thanks{\texttt{stariate@ee.mount-sheikh.edu}}}%
}
\affil[1]{Department of Computer Science, Cranberry-Lemon University, Pittsburgh, PA 15213}
\affil[2]{Department of Electrical Engineering, Mount-Sheikh University, Santa Narimana, Levand}
\fi

\renewcommand{\undertitle}{Submitted to 19th International Conference on the Design of Reliable Communication Networks (DRCN)}
\renewcommand{\shorttitle}{}

\hypersetup{
pdftitle={Optimum Network Slicing for Ultra-reliable Low Latency Communication (URLLC)},
pdfsubject={Computer Science},
pdfauthor={Iulisloi Zacarias, Francisco Carpio, André Costa Drummond, Admela Jukan},
pdfkeywords={},
}

\begin{acronym}
\acro{4G}{fourth generation of cellular networks}
\acro{5G}{fifth generation of cellular networks}

\acro{BBU}{base band unit}

\acro{CapEx}{capital expenditure}
\acro{CN}{core network}
\acro{C-RAN}{centralized RAN}
\acro{CS}{cell site}
\acro{CU}{centralized unit}

\acro{D-RAN}{distributed RAN}
\acro{DN}{data network}
\acro{DU}{distributed unit}

\acro{E2E}{end-to-end}
\acro{eMBB}{enhanced mobile broadband}
\acro{eNB}{evolved node B}

\acro{GOPS}{giga operations per second}

\acro{ILP}{integer linear programming}
\acro{IoT}{internet of things}
\acro{InP}{infrastructure provider}

\acro{LP}{linear programming}

\acro{MEC}{multi-access edge computing}
\acro{MILP}{mixed integer linear programming}
\acro{mMTC}{massive machine-type communication}

\acro{NF}{network function}
\acro{NG-RAN}{next-generation RAN}
\acro{NPN}{non-public 5G network}
\acro{NS}{network slicing}
\acro{NSI}{network slice instance}

\acro{OpEx}{operationalexpenditures}

\acro{QoS}{quality of service}

\acro{RAN}{radio access network}
\acro{RF}{radio frequency}
\acro{RRH}{remote radio head}
\acro{RU}{radio unit}

\acro{SLA}{service-level agreement}

\acro{TN}{transport network}

\acro{UE}{user equipment}
\acro{URLLC}{ultra reliable low latency communication}
\acro{UPF}{user plane function}

\acro{VNF}{virtual network function}
\acro{VM}{virtual machine}

\end{acronym}
\begin{document}
\maketitle

\begin{abstract}
	Within 3GPP, the campus network architecture has evolved as a deployment option for industries and can be provisioned using network slicing over already installed 5G public network infrastructure. In campus networks, the ultra-reliable low latency communication (URLLC) service category is of major interest for applications with strict latency and high-reliability requirements. One way to achieve high reliability in a shared infrastructure is through resource isolation, whereby network slicing can be optimized to adequately reserve computation and transmission capacity. This paper proposes an approach for vertical slicing the radio access network (RAN) to enable the deployment of multiple and isolated campus networks to accommodate URLLC services. To this end, we model RAN function placement as a mixed integer linear programming problem with URLLC-related constraints. We demonstrate that our approach can find optimal solutions in real-world scenarios. Furthermore, unlike existing solutions, our model considers the user traffic flow from a known source node on the network's edge to an unknown \textit{a priori} destination node. This flexibility could be explored in industrial campus networks by allowing dynamic placement of user plane functions (UPFs) to serve the URLLC.
\end{abstract}

\section{Introduction}

The \ac{5G} has fostered innovation in the industrial world, making it possible to transmit data at very low latencies and high reliability. The so-called vertical slicing of \ac{RAN} is essential to creating public-network integrated campus networks, which are especially attractive in Industry 4.0~\cite{5g-acia-npn}. To this end, the base stations can be split into the three functional blocks: \ac{RU}, \ac{DU}, and \ac{CU}. With virtualization and \textit{softwarization} of these radio functions, we can create dedicated and isolated \ac{RAN}, which are, in effect, the provisioned campus network~\cite{Morais2022}. To guarantee the reliability, the network operator must ensure proper slice resource provisioning while dealing with failures or varying service requirements \cite{Gomes2020}. To this end,  the redundancy of resources and the isolation between slices is essential to guarantee steady performance. 

This work investigates the efficient usage of \ac{NS} to provision dedicated \ac{RAN} for campus networks requiring high isolation for \ac{URLLC} type of industrial environments. We consider a fully disaggregated RAN composed of \ac{RU}, virtual \ac{CU}, and virtual \ac{DU}, whereby each functional block can be placed individually in the network~\cite{Morais2022}. The proposed approach addresses the use of \ac{RAN} slices in a logical isolation setup by deploying dedicated \acp{NF} for each slice~\cite{Yu2020}. As a deployment scenario, a multi-tiered network is considered. The radio functions can be placed in small data centers at the network's edge, aiming for low latency (i.e., a \ac{URLLC} slice). The core cloud can be explored for services with less stringent requirements, where computing resources are abundant and the computation delays are reduced. The main idea is to make the least strict applications to be computed closer to the core while reserving edge computation availability for \ac{URLLC}. Efficient algorithmic solutions to optimize the slice allocation, as the one addressed in this paper, are still an open challenge towards feasible \ac{URLLC} services.

We formulate the problem as a \ac{MILP} aiming to find the best trade-off between the average computation load of servers and the traffic load of logical links while guaranteeing the requirements of \ac{URLLC} services. We demonstrate that our approach can find optimal solutions in real-world scenarios. Unlike existing solutions, our model considers the user traffic flow from a known source node on the network's edge to an unknown \textit{a priori} destination node since its location will depend on where the \ac{CU} is placed. This flexibility could be explored in industrial campus networks by allowing dynamic placement of \acp{UPF} to meet the \ac{URLLC} requirements. Our novel contribution is also in joint consideration of computation and networking delays.

The rest of this paper is organized as follows: Section~\ref{sec:rw} presents the related work. Section~\ref{sec:cn-slicing} describes the reference network and the optimization model. The scenario evaluated is presented in Section~\ref{sec:scenario}. Section~\ref{sec:results} presents the results. Section~\ref{sec:conclusion} concludes the paper.

\section{Related Work}
\label{sec:rw}

When implementing campus networks with network slicing, efficient placement of \ac{RAN} functions plays an essential role. Disaggregated \ac{5G} \ac{RAN} function placement using general purpose hardware is investigated in \cite{Morais2022}, \cite{Yu2020}, \cite{Sarikaya2021}, \cite{Ojaghi2022}, and \cite{Coelho2022}. The authors investigated the problem from different perspectives. For example, \cite{Yu2020} aim at energy efficiency when placing RAN functions. \cite{Ojaghi2022} focus on minimizing the economic cost of \ac{RAN} and \ac{MEC} deployments while maximizing the served traffic. \cite{Morais2022} finds the trade-off between maximum aggregation level (\ie maximum amount of grouped \ac{RAN} \acp{NF}) and minimum computing resources. 

Our approach is similar to \cite{Morais2022}, \cite{Yu2020} and \cite{Sarikaya2021}, which consider a multi-tier network with varying computing resources at each tier. The slicing problem of \ac{RAN} is tackled in~\cite{Sarikaya2021}, \cite{Ojaghi2022}, and~\cite{Coelho2022}. Although the authors consider different types of network slices (\ac{URLLC}, \ac{eMBB}, and \ac{mMTC}) and their varying and sometimes contrasting requirements regarding network capacity and latency, a fixed latency is attributed to queueing, and processing of traffic by \ac{RAN} functions. Since the requirements for the same type of slices can vary to some extent (\eg \ac{URLLC} slices with different bandwidth requirements), these slices can experience different \ac{E2E} service delays affecting the final \ac{QoS} of the slice. 

To contribute to the existing body of knowledge, this work applies \ac{MILP} to solve the problem of vertical slicing \ac{RAN} by allowing the sharing of public infrastructure among slices with strict and, at times, competing requirements (e.g., \ac{URLLC} and \ac{eMBB} slices sharing the same \ac{RAN}). Our approach solves this problem in real-world multi-tiered topologies, just like the previously mentioned work. However, unlike previous work, we consider the delays introduced by processing and queuing at the servers in addition to network delays, which, depending on the size of the demand, can largely impact the total delay experienced by the service. In this way, we consider the delay caused by the processing of one slice to other slices sharing the same server, guaranteeing that the traffic load applied to one slice does not cause \ac{SLA} violations in other slices, which is critical to \ac{URLLC} performance.

\section{Campus Network Slicing}
\label{sec:cn-slicing}

This section presents the reference industrial campus network scenario and the characterization of applications supported in typical industrial premises. After that, the formal system optimization model is presented.

\subsection{Reference Network Scenario}

We consider a scenario in which a campus network is deployed for an industry vertical hosted by the public network, i.e., public-network integrated deployment~\cite{5g-acia-npn}. The public mobile network operator uses network slicing to deploy the campus network according to the requirements, including reliability and security.

\begin{figure}[htbp]
\centerline{\includegraphics[width=0.5\columnwidth]{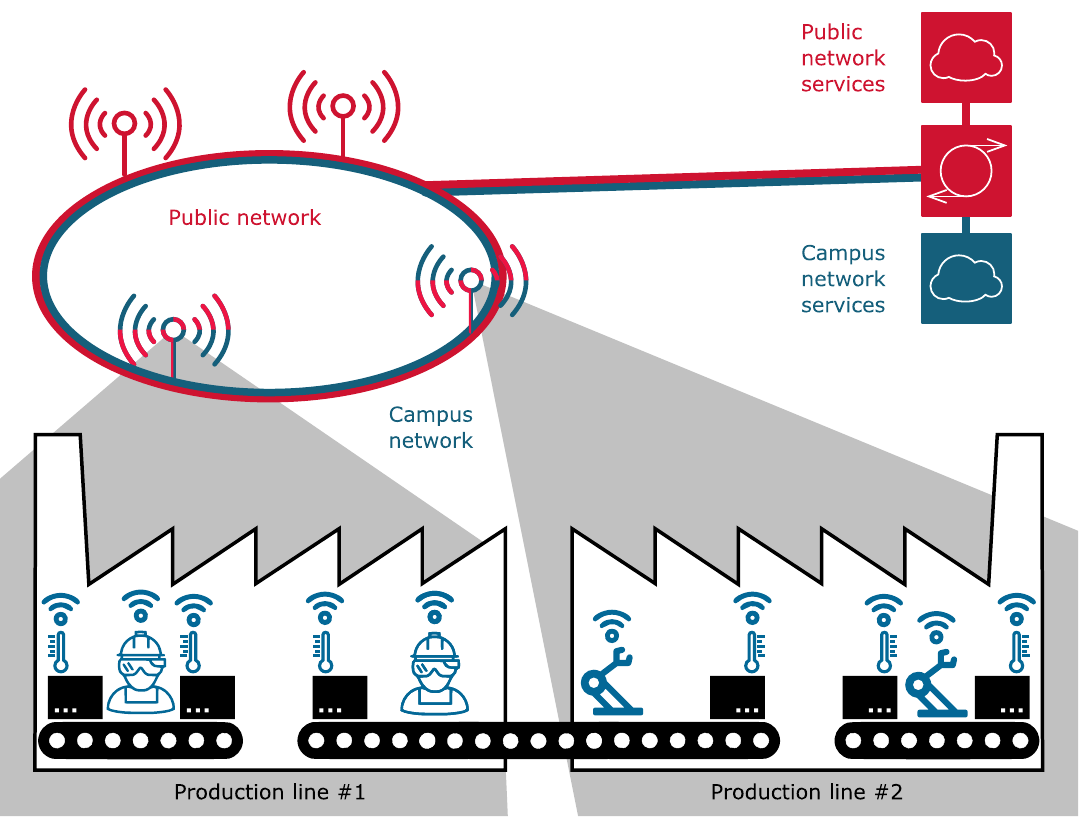}}
\caption{Deployment scenario of a public network integrated campus (non-public) network in a factory plant composed of several production lines connected by multiple base stations. Adapted from~\cite{5g-acia-npn}.}
\label{fig:pni-npn-network}
\end{figure}

\figref{fig:pni-npn-network} illustrates the scenario. We assume that a single campus network, composed of two Production lines connected by separate \acp{BBU}, can require different slices depending on its applications' requirements. In Production line \#1, a slice of type \ac{URLLC} is required to provide connectivity to applications supporting the augmented workers. This application must provide real-time, high-quality video rendering while reacting to the worker's head movements. Thus it requires low latency, high reliability, and high bandwidth. The second slice of type \ac{URLLC} is required to connect the autonomous robots in Production line \#2. This application also requires low latency and high reliability to ensure the correct functioning of the robot safely, however with less stringent requirements in terms of bandwidth. Moreover, both Production lines need one more slice each for the equipment temperature monitoring service, being those slices of type \ac{mMTC}. Finally, this campus network comprises four 5G public network slices.

Since the slices composing the campus network share the \ac{RAN} with the public network and other slices, vertical network slicing is employed to guarantee the requirements of the \ac{URLLC} slices. Furthermore, by employing the virtualization of the \acp{NF}, the mobile network operator can provide a dedicated \ac{RAN} for each slice, ensuring the isolation among slices with competing requirements or among slices belonging to different industries. The slice isolation is achieved by providing a dedicated radio protocol stack employing virtualization of \ac{RAN} functional blocks for each application.

The \ac{NG-RAN} proposed for 5G implements a decoupled radio protocol stack allowing the splitting of base stations into up to three functional blocks: \ac{RU}, \ac{DU}, and \ac{CU}. Additionally, softwarization and virtualization of radio functions can reduce the cost of \ac{RAN} deployments by placing each functional block individually in the virtualized network on top of general-purpose hardware~\cite{Morais2022}.

\begin{figure}[htbp]
    \centerline{\includegraphics[width=0.5\columnwidth]{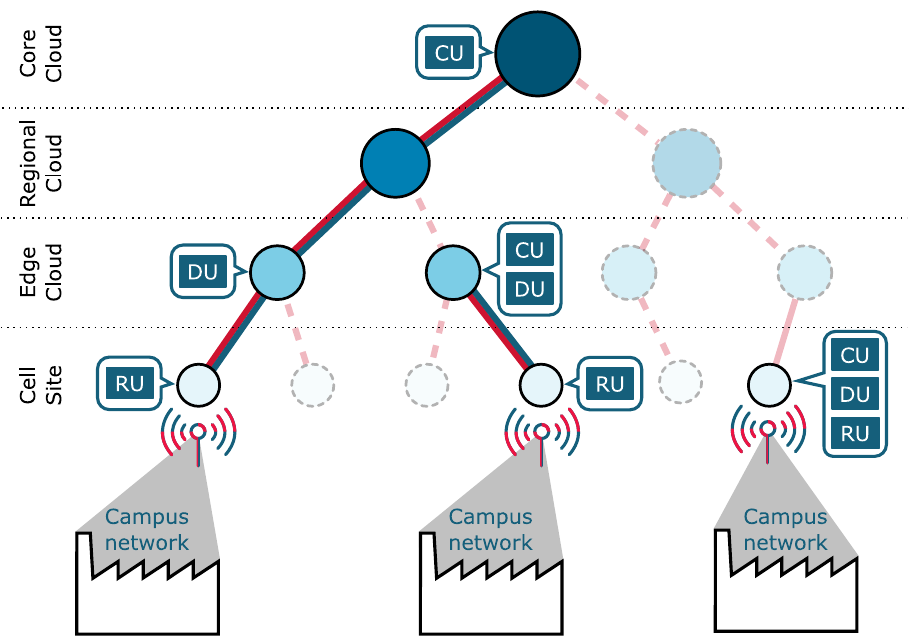}}
    \caption{Three examples of virtual \acp{RAN} allocation in a multi-tier network.}
    \label{fig:npn-topo}
\end{figure}

The campus network services, represented in \figref{fig:pni-npn-network} as a blue cloud box, are provided by the instantiation of a set of isolated virtual \acp{RAN}, each composed of a separate \ac{NF} chain (\ac{RU}, \ac{DU}, and \ac{CU}). \figref{fig:npn-topo} shows examples of possible allocations of \acp{NF} to provide virtual \acp{RAN} in a multi-tier network with computational nodes layered according to its proximity to the \acp{BBU}. The \ac{CS} is the closest location where a \ac{NF} can be placed hosting \acp{RU} providing radio interface to \acp{UE}, but also the one with the least amount of computational resources. On top of that, three levels of cloud computing nodes can be considered, with the core cloud having the larger computational capacity. On the other hand, there are far more instances of \acp{CS} compared to core cloud sites. Three different slices are presented as part of separate campus networks. The slices' configuration differs from each other considering link and computation availability, resource cost, and the service requirements; \eg an \ac{URLLC} slice will need a shorter \ac{NF} chain, with \ac{DU} and \ac{CU} allocated closely to the \ac{BBU}. Therefore, the closer the \ac{NF} to the \ac{BBU}, the lower the network delay. At the same time, nodes closer to the edge have fewer computing resources, which leads to higher computational delays. Therefore, a trade-off between network and computational resource allocation needs to be addressed to meet the requirements of \ac{URLLC} services.

\subsection{System Model}
\label{sec:system-model}


The network is modeled as a graph $\mathbb{G} = (\mathbb{N} \cup \mathbb{X}, \mathbb{L})$ where $\mathbb{N} = \{1, \ldots, N\}$ is a set of sites hosting a set of servers $\mathbb{X} = \{1, \ldots, X \}$ and a set $\mathbb{L} = \{1, \ldots L \}$ of directed links connecting the sites. The delay of connection among servers placed on a site (i.e., internal data center connections) is disregarded in this work. Each link $ l \in \mathbb{L}$ connecting two sites has limited capacity regarding bandwidth (given in Mbps) and also a propagation delay (given in fraction of a second), represented by $C_{l}^{max}$ and $D_{l}$ respectively. The set $ \mathbb{S} = \{1, \ldots, S \}$ denotes all RAN slice instances provisioned in the network, which can be implemented as a collection of \acp{NF} composing a virtualized network service~\cite{Morais2022}. Therefore, a specific slice instance $ s \in \mathbb{S}$ corresponds to an ordered set of \acp{NF} denoted by $ \mathbb{V}_{s} = \{1, \ldots, V_{s}\}$ where the \ac{NF} $ v \in \mathbb{V}$ is the $v^{\text{th}}$ \ac{NF} in the chain composing the \ac{RAN} slice. Each \ac{NF} $ v \in \mathbb{V} $ has its type $ t \in \mathbb{T} $, which defines the parts of the radio protocol stack that the \ac{NF} supports. Since \acp{NF} are not shared among slices, we guarantee the isolation of each slice, even when sharing the same server. Without loss of generalization, we assume that each campus network is composed of a subset of one or more slices $s \in \mathbb{S}$.

The traffic demands were modeled having the \ac{CS} as its source, and the last \ac{NF} in the set $ \mathbb{V}_{s} $ (\ie the CU) as its destination. Hence the destination node of the traffic demand can vary, since the placement of the \ac{CU} depends on the model decision. We assume that each \ac{UE} is connected to a single \ac{RU} placed at one of the \ac{CS}. The input set $ p \in \mathbb{P}_{s}$ is the set of k-shortest paths from the source node in the edge to its corresponding \ac{CU} that can be placed in any intermediate node between the \ac{CS} and the Core Cloud sites (including these sites). All parameters considered in the problem formulation are summarized in \tabref{tb:parameters}, while the variables used in the \ac{MILP} model are summarized in \tabref{tb:variables}.

\begin{table}[htb]
\renewcommand{\arraystretch}{1.1}
\caption{List of parameters used in the model}
\centering
\scriptsize
\begin{tabular}{p{1.1cm}l}
\toprule
\textbf{Parameter} & \textbf{Meaning} \\ 
\midrule
$\mathbb{N}$                    & set of sites: $N = \{1, \ldots, N\}, n \in N$. \\ 
$\mathbb{X}$                    & set of servers: $X = \{1, \ldots, X\}, x \in X$. \\ 
$\mathbb{L}$                    & set of links: $L = \{1, \ldots, L\}, l \in L$.   \\ 
$\mathbb{P}$                    & set of admissible paths: $P = \{1, \ldots, P\}, p \in P$. \\ 
$\mathbb{S}$                    & set of slices: $S = \{1, \ldots, S\}, s \in S$. \\ 
$\mathbb{T}$                    & set of NF types: $T = \{ \text{RU}, \text{DU}, \text{CU} \}$. \\ 
$\mathbb{V}_{s}$                & ordered set of NFs, $V_s  = \{ \text{RU}_s, \text{DU}_s, \text{CU}_s \}$. \\ 
$\mathbbl{\Lambda}$             & set of traffic demands: $\mathbbl{\Lambda} = \{\lambda, \ldots, \mathbbl{\Lambda}\}$. \\ 
$\mathbbl{\Lambda}_{s} \subset \mathbbl{\Lambda}$   & subset of traffic demands $\lambda \in \mathbbl{\Lambda}$ for slice $s \in S$. \\ 
$\mathbb{N}_p \subset \mathbb{N}$     & subset of ordered sites in path $p \in P$. \\
$\mathbb{X}_{n} \subset \mathbb{X}$   & subset of servers attached to site $N$. \\ 
$\mathbb{X}_{p} \subset \mathbb{X}$   & ordered subset of servers composing path $p$. \\ 
$\mathbb{P}_{s} \subset \mathbb{P}$   & subset of admissible path $p \in P$, for $s \in S$. \\ 
$T_{p}^{l}$                     & binary, 1 if path $p \in P$ traverses link $l \in L$. \\ 
$T_{p}^{n,m}$                   & binary, 1 if path $p \in P$ connects node  $n \in N$ and $m \in N$. \\ 
$\Gamma_{t(v)}^{pro}$           & continuous, load ratio of a NF $v \in V$ of type $t \in T$. \\ 
$C_{l}^{max}$                   & integer, maximum capacity of the link $ l \in L$. \\ 
$C_{x}^{max}$                   & integer, maximum capacity of server $x \in X$. \\ 
$C_{x, t(v)}^{\text{proq, max}}$ & integer, maximum processing capacity that can be assigned by \\ 
                                 & a server $x$ to a NF of type $t$ \\
$D_{l}$                         & continuous, propagation delay of link $l \in L$. \\ 
$D_{s}^{max}$                   & continuous, maximum service delay of a slice $s \in S$. \\ 
$D_{t(v)}^{pro,max}$            & continuous, maximum allowed processing delay  for a NF of type $t$. \\ 
$D_{t(v)}^{pro,min}$            & continuous, minimum processing delay for a NF of type $t$. \\ 
$D_{t(v)}^{proq}$               & continuous, delay of a NF $v$ of type $t$ due to queueing. \\ 
$D_{t(v)}^{prox}$               & continuous, delay of a NF $v$ of type $t$ due to processing. \\
\bottomrule
\end{tabular}
\label{tb:parameters}
\renewcommand{\arraystretch}{1}
\end{table}

\begin{table}[htbp]
\renewcommand{\arraystretch}{1.1}
\caption{List of variables used in the model}
\centering
\scriptsize
\begin{tabular}{p{0.85cm}l}
\toprule
\textbf{Variable}               & \textbf{Meaning} \\ 
\midrule
$z_{p}^{s}$                     & binary, 1 if slice $s$ uses path $p \in P_{s}$. \\ 
$z_{p}^{\lambda, s}$            & binary, 1 if traffic demand $\lambda$ from SFC $s$ uses path $p \in P_{s}$. \\ 
$f_{x}$                         & binary, 1 if server $x$ is used. \\ 
$f_{x}^{v,s}$                   & binary, 1 if NF $v$ from SFC $s$ is allocated at server $x$. \\ 
$f_{x,\lambda}^{v,s}$           & binary, 1 if NF $v$ from SFC $s$ is used at server $x$ by traffic demand $\lambda$. \\
$d_{p}^{\lambda,s}$             & continuous, service delay of a traffic demand $\lambda$ in a path $p$. \\
$k_{x}$                         & continuous, utilization cost of server $x \in X$. \\ 
$k_{l}$                         & continuous, utilization cost of a link $l \in L$. \\ 
$u_{x}$                         & continuous, utilization of server $x \in X$. \\ 
$u_{l}$                         & continuous, utilization of a link $l \in L$. \\ 
\bottomrule
\end{tabular}
\label{tb:variables}
\renewcommand{\arraystretch}{1}
\end{table}

\subsection{Objective function}

The problem formulation has the objective of minimizing the cost of all servers and links in the network, \ie
\begin{equation*}
    {minimize}: 
    \frac{\alpha}{\vert \mathbb{X}\vert }\sum_{x\in \mathbb{X}}k_{x} + 
    \frac{1 - \alpha}{\vert \mathbb{L} \vert }\sum_{\ell\in \mathbb{L}}k_{\ell}
\end{equation*}
where $k_{x}$ is the utilization cost of servers and $k_{\ell}$ is the utilization cost of links. The $\alpha$ parameter allows us to select the desired trade-off between the server and link cost utilization. The utilization cost of a server is defined in \eqref{eq:server-cost}, where the variable $f_{x, \lambda}^{v,s}$ specifies if a \ac{NF} $v$ uses the server $x$ for processing a demand $\lambda \in \mathbbl{\Lambda}_s$ from a slice $s$. If the variable is 1,  the traffic from $\lambda$, multiplied by the corresponding load ratio of \ac{NF} $v$ is added to the server utilization. The cost is normalized by the maximum capacity of server $x$. 

The link utilization cost is defined in \eqref{eq:link-cost}. The variable $z_{p}^{\lambda, s}$ defines whether a traffic demand $\lambda$ from a slice $s$ is using path $p$ and $T_{p}^{l}$ checks if path $p$ contains link $l$ in order to sum the traffic demand from $l$ to the link utilization. Similarly to \eqref{eq:server-cost}, the link utilization is normalized by dividing the current link utilization by the total capacity of the link. In both equations \eqref{eq:server-cost} and \eqref{eq:link-cost}, $a_{i}$ and $b_{i}$ are terms of a piecewise linearization function.


\begin{multline}
   \label{eq:server-cost}
    k_{x} = 
    a_{i} \Bigg[
    \frac{1}{C_{x}^{\text{max}}}
    \sum_{s \in \mathbb{S}}
    \sum_{v \in \mathbb{V}_{s}}
    \left( 
        \Gamma_{t\left( v \right)}^{\text{pro}}
        \sum_{\lambda \in \mathbbl{\Lambda}_{s}}
        \lambda \cdot f_{x,\lambda}^{v,s} 
    \right)
    \Bigg]
    - b_{i},
    \forall x \in \mathbb{X}
    ~, 
\end{multline}

\begin{multline}
\label{eq:link-cost}
    k_{\ell} = 
    a_{i} \Bigg[
    \frac{1}{C_{\ell}^{\text{max}}}
    \sum_{s \in \mathbb{S}}
    \sum_{p \in \mathbb{P}_{s}}
    \sum_{\lambda \in \mathbbl{\Lambda}_{s}}
    \lambda \cdot T_{p}^{\ell} \cdot z_{p}^{\lambda,s}
    \Bigg]
    - b_{i},
    \forall \ell \in \mathbb{L}
\end{multline}
subject to a set of constraints that are described next.

\subsection{Constraints}
The following two equations enable the routing of demands from the \ac{CS} to the destination \ac{CU} node, which can be placed in any node of the path. For each demand $\lambda \in \mathbbl{\Lambda}_s$ of slice $s$, exactly one path $p \in \mathbb{P}_{s}$ must be selected, as represented by the constraint~\eqref{eq:rp1}. The binary variable $z_{p}^{\lambda, s}$ indicates that a traffic demand $\lambda \in \mathbbl{\Lambda}_{s}$ of slice $s$ is using the path $p \in \mathbb{P}_s$. Equation \eqref{eq:rp2} takes the activated path from $z_{p}^{\lambda, s}$ and activates the same path for the corresponding slice instance. The left side forces $z_{p}^{s}$ to be 1 when at least one demand of slice $s$ is using path $p$ while the right part forces $z_{p}^{s} = 0$ when no demands of slice $s$ are using path $p$.

\begin{equation}
    \label{eq:rp1}
    \sum_{p \in \mathbb{P}_{s}}
    z_{p}^{\lambda, s} = 1, 
    \forall s \in \mathbb{S}, ~
    \forall \lambda \in \mathbbl{\Lambda}_{s}
\end{equation}

\begin{equation}
    \label{eq:rp2}
    z_{p}^{\lambda, s} \le z_{p}^{s} \le 
    \sum_{\lambda^{\prime} \in \mathbbl{\Lambda}_{s}} z_{p}^{\lambda^{\prime},s}, 
    \forall s \in \mathbb{S},
    \forall p \in \mathbb{P}_{s},
    \forall \lambda \in \mathbbl{\Lambda}_{s}
\end{equation}

The next three constraints are responsible for activating \acp{NF} along the path and mapping traffic demands to \acp{DU} and \acp{CU} in an isolated fashion, for providing higher reliability. Equation~\eqref{eq:pf1} assures that each traffic demand from a slice is processed by every \ac{NF} (\ie \ac{RU}, \ac{DU}, \ac{CU}) in one specific server. The binary variable $f_{x,\lambda}^{v,s}$ assumes value 1 if traffic demand $\lambda \in \mathbbl{\Lambda}_s$ is using a \ac{NF} $v \in V_s$ from slice $s$ at server $x \in \mathbb{X}$. 

\begin{equation}
    \label{eq:pf1}
    \sum_{x \in \mathbb{X}} f_{x,\lambda}^{v,s} = 1, ~
    \forall s \in \mathbb{S},
    \forall v \in \mathbb{V}_{s},
    \forall \lambda \in \mathbbl{\Lambda}_{s}
\end{equation}

The constraint \eqref{eq:pf2} takes the activated \ac{NF} for each demand from the previous equation and activates the \ac{NF} for the slice that the demand belongs to. The left side of the inequality forces $f_{x}^{v,s}$ to assume the value 1 when at least one traffic demand $\lambda \in \mathbbl{\Lambda}_s$ is using a \ac{NF} $v \in \mathbb{V}_s$ at server $x \in \mathbb{X}$ whilst the right side forces $f_{x}^{v,s}$ to be 0 when no traffic demand $\lambda \in \mathbbl{\Lambda}_s$ is using that specific \ac{NF} $v \in \mathbb{V}_s$ at a specific server $x \in \mathbb{X}$. 

\begin{equation}
    \label{eq:pf2}
    f_{x,\lambda}^{v,s} \le
    f_{x}^{v,s} \le 
    \sum_{\lambda^{\prime} \in \mathbbl{\Lambda}_{s}} f_{x, \lambda^{\prime}}^{v,s},
    \forall s \in \mathbb{S},
    \forall v \in \mathbb{V}_{s},
    \forall x \in \mathbb{X},
    \forall \lambda \in \mathbbl{\Lambda}_{s}
\end{equation}

To determine wheter a server is used, the binary variable $ f_{x}$ is constrained according to \eqref{eq:pf3}. The left side of the inequality sets the value of $f_{x}$ to 1 when at least one \ac{NF} $v$ of any slice $s \in \mathbb{S}$ is allocated at server $x \in \mathbb{X}$ while the right part forces $f_{x}$ to be 0 if the server is not being used by any \ac{NF}.

\begin{multline}
    \label{eq:pf3}
    \frac{1}{|\mathbb{S}||\mathbb{V}_{s}|}
    \sum_{s \in \mathbb{S}} \sum_{v \in \mathbb{V}_{s}} f_{x}^{v,s} 
    \leq f_{x} \leq
    \sum_{s \in \mathbb{S}} \sum_{v \in \mathbb{V}_{s}} f_{x}^{v,s}, ~
    \forall x \in \mathbb{X}
\end{multline}

The number of instances of \ac{NF} $v \in \mathbb{V}_{s}$ belonging to a slice $ s \in \mathbb{S} $ that can be activated in a server is limited by the constraint described in~\eqref{eq:fd1}. This constraint limits the number of instances of a specific \ac{NF} $v \in \mathbb{V}_s$ from slice $s\in \mathbb{S}$ to 1, allowing a maximum of one instance of $\mathbb{V}_s= \{ \text{RU}, \text{DU}, \text{CU} \}$ being allocated for each slice. 

\begin{equation}
    \label{eq:fd1}
    \sum_{x \in \mathbb{X}} f_{x}^{v,s} \leq 1, 
    \forall s \in \mathbb{S},
    \forall v \in \mathbb{V}_{s}.
\end{equation}

Constraint~\eqref{eq:fd2} maps all the \acp{NF} in the set $ \mathbb{V}_{s} $ on the activated path for slice $s \in \mathbb{S} $ forcing every \ac{NF} $v \in \mathbb{V}_{s}$ to be instantiated in any server $x \in \mathbb{X}_{p}$ placed along the path $p \in \mathbb{P}_{s}$ for the demands $\lambda \in \mathbbl{\Lambda}_{s}$. When $z_{\lambda}^{s, p}$ is 1, at least one instance of each specific \ac{NF} $v \in \mathbb{V}_s= \{ \text{RU}, \text{DU}, \text{CU} \}$ must be activated. On the other hand, when $z_{\lambda}^{s, p}$ is 0, no \acp{NF} can be allocated on servers placed along the path. 

\begin{equation}
    \label{eq:fd2}
    \sum_{x \in \mathbb{X}_{p}} 
    f_{v,s}^{x, \lambda}
    \geq z_{p}^{\lambda,s}, 
    \forall s \in \mathbb{S},
    \forall p \in \mathbb{P},
    \forall \lambda \in \mathbbl{\Lambda}_{s},
    \forall v \in \mathbb{V}_{s}
\end{equation}

Because each slice is composed of an ordered set of \acp{NF} (\ie RU, DU, CU), the traffic demand has to traverse all these \acp{NF} in the correct order enforced by constraint~\eqref{eq:fd3}. The left part of the inequation guarantees the ordering of \acp{NF} and it is activated only when $z_{p}^{\lambda,s}$ is 1, meaning that path $p \in \mathbb{P}_{s}$ is activated. Therefore, for each traffic demand $\lambda$ of slice $s$, the $v^{\text{th}}$ \ac{NF} is only allocated at server $x \in \mathbb{X}_n$ if the previous \ac{NF} $v^{\text{th} - 1}$ is allocated in any server $y \in \mathbb{X}_m$, where $m$ ranges from 1 to the $n$ traversed site in path $p$. The ordering of servers $x \in \mathbb{X}_n$ in site $n$ is not modeled in our approach, and we assume that local routing policies will enforce the correct ordering inside these subsets. 

\begin{multline}
    \label{eq:fd3}
    \left(\sum_{m=1}^n \sum_{y \in \mathbb{X}_{m}} f_{y, \lambda}^{(v-1), s} \right) - 
    \sum_{x \in \mathbb{X}_n} f_{x,\lambda}^{v,s} \geq
    z_{p}^{\lambda,s} - 1
    \begin{cases}
    \begin{matrix}
    1 < v \le |\mathbb{V}_{s}| \\
    n \ne m
    \end{matrix}~, 
    \end{cases}\\
    \forall s \in \mathbb{S},
    \forall \lambda \in \mathbbl{\Lambda}_{s},
    \forall p \in \mathbb{P}_{s},
    \forall v \in \mathbb{V}_{s},
    \forall m,n \in \mathbb{N}
\end{multline}

Since RUs run part of \ac{RF} protocol, they should always be placed close to the \ac{RF} antennas and the user equipment. The constraint~\eqref{eq:src} assures their placement at the first site composing the path, therefore always placing these type of \acp{NF} at the \ac{CS}.

\begin{equation}
    \label{eq:src}
    \sum_{x \in \mathbb{X}_{p}} f_{x}^{v,s} = 1,~\forall s \in \mathbb{S}, n = 0, v = 0, p = 0.
\end{equation}

The parameter $D_{s}^{\text{max}}$ specifies the total service delay of a slice. Therefore, the \ac{E2E} user plane delay from the \ac{RU} to the \ac{CU} is considered, being the placement of the \ac{CU} dependent on the model's decision. The \ac{E2E} service delay depends on the propagation delay of traffic demands when traversing links, plus the processing delay of \acp{NF} in servers, given by \eqref{eq:delay1}. In \eqref{eq:delay2}, the total processing load of a \ac{NF} in the server $x$ is controlled by the binary variable $f_{x, \lambda}^{v, s}$. Hence, if $f_{x, \lambda}^{v, s}$ is 1, a fraction of the queueing delay $D_{t(v)}^{\text{proq}}$ is accounted. The delay term given in \eqref{eq:delay3} adds the load independent minimum delay of a \ac{NF} depending on its type $t \in \mathbb{T}$ and a varying delay part that linearly increases with the server utilization, which is calculated acording to \eqref{eq:server-util}.

\begin{subequations}
    \begin{align}
    d_{x, v, s}^{\text{pro}} &= d_{x, v, s}^{\text{proq}} + d_{x, v, s}^{\text{prox}}, 
    \forall s \in \mathbb{S}, \forall v \in \mathbb{V}_s, \forall x \in \mathbb{X}_p \label{eq:delay1} \\%
    d_{x, v, s}^{\text{proq}} &= D_{t,v}^{proq} \frac{\Gamma_{t(v)}^{pro} \cdot \sum_{\lambda \in \mathbb{\Lambda}_s} f_{x,\lambda}^{v,s}}{C_{x, t(v)}^{\text{proq, max}}}  \label{eq:delay2} \\%
    d_{x, v, s}^{\text{prox}} &= D_{t(v)}^{pro,min} \cdot F_{x}^{v,s} + D_{t(v)}^{prox} \cdot u_{x} \label{eq:delay3}
    \end{align}  
\end{subequations}

\begin{equation}
    \label{eq:server-util}
     u_{x}  = 
     \frac{1}{C_{x}^{\text{max}}}
     \sum_{s \in \mathbb{S}}
     \sum_{v \in \mathbb{V}_{s}}
     \left( 
         \Gamma_{t\left( v \right)}^{\text{pro}}
         \sum_{\lambda \in \mathbbl{\Lambda}_{s}}
         \lambda \cdot f_{x,\lambda}^{v,s} 
     \right),
     \forall x \in \mathbb{X}
     ~, 
 \end{equation}

 Finaly, the total \ac{E2E} delay of the slice on the selected path $p \in \mathbb{P}_s$ is given in \eqref{eq:max-delay} and it is constrained by the maximum service delay of the slice $D_{s}^{\text{max}}$. The first term is the propagation delay of demands on links, which is controlled by $T_{p}^{\ell}$ and it is 1 only when the link is used by path $p$. The second term add the processing delay of the \ac{NF} in a server $x \in \mathbb{X}_p$, which is controlled by the variable $f_{x,\lambda}^{v,s}$. 

\begin{equation}
    \begin{split}
    \label{eq:max-delay}
    \sum_{\ell \in \mathbb{L}} D_{\ell} \cdot T_{p}^{\ell} + 
    \sum_{x \in \mathbb{X}_{p}} \sum_{v \in \mathbb{V}_{s}} d_{x,v,s}^{\text{pro}}(\vec{\lambda}) \cdot f_{x,\lambda}^{v,s}
    \leq D_{s}^{\text{max}}\\
    \forall s \in \mathbb{S}, \forall \lambda \in \mathbbl{\Lambda}_{s}, \forall p \in \mathbb{P}_{s}
    \end{split}
\end{equation}

\subsection{First-fit algorithm (heuristics)}

A first-fit algorithm (heuritics) is proposed for comparison purposes. Based on the O-RAN specifications~\cite{ORAN-WG6-CAD}, the algorithm realizes the function placement by ordering the set of all slice requests by the required \ac{E2E} delay latency, starting with the lowest first. Sequentially, for each slice demand, the set of all admissible paths (paths containing the source and destination nodes of the demand) is selected. Next, the paths are filtered depending on the slice service type. For slices of type \ac{URLLC}, only paths containing nodes with available servers at the \ac{CS} or Edge Cloud are selected. For slices of type \ac{eMBB}, first, we try to select paths that contain nodes with servers available at the Regional Cloud. If no servers are available, paths with servers at the Edge Cloud are selected. If still no servers are available, paths containing sites and servers at the CS are selected. For slices of type mMTC, paths containing servers at the Core Cloud are selected. In the final step, each RAN block composing the slice is placed in the first available server of the first path in the set of paths.

The proposed algorithm does not consider the maximum \ac{E2E} service delay of the slice to allocate the RAN blocks. Because each new \ac{NF} allocated at a server changes the processing delay of the already allocated \acp{NF} at that server, such a feature requires a complete state space search, what was already done by \ac{MILP} model. The time complexity of the first-fit algorithm is $O(|\mathbb{S}| ((|\mathbbl{\Lambda}_{s}| \cdot |\mathbb{P}_{s}|) + |\mathbb{V}_{s}|))$. Note that $\mathbb{P}_{s}$ and $\mathbb{V}_{s}$ are sets with small cardinality regardless of the network scenario.

\section{Evaluation Scenario}
\label{sec:scenario}

\begin{figure}[htbp]
\centerline{\includegraphics[width=0.5\columnwidth]{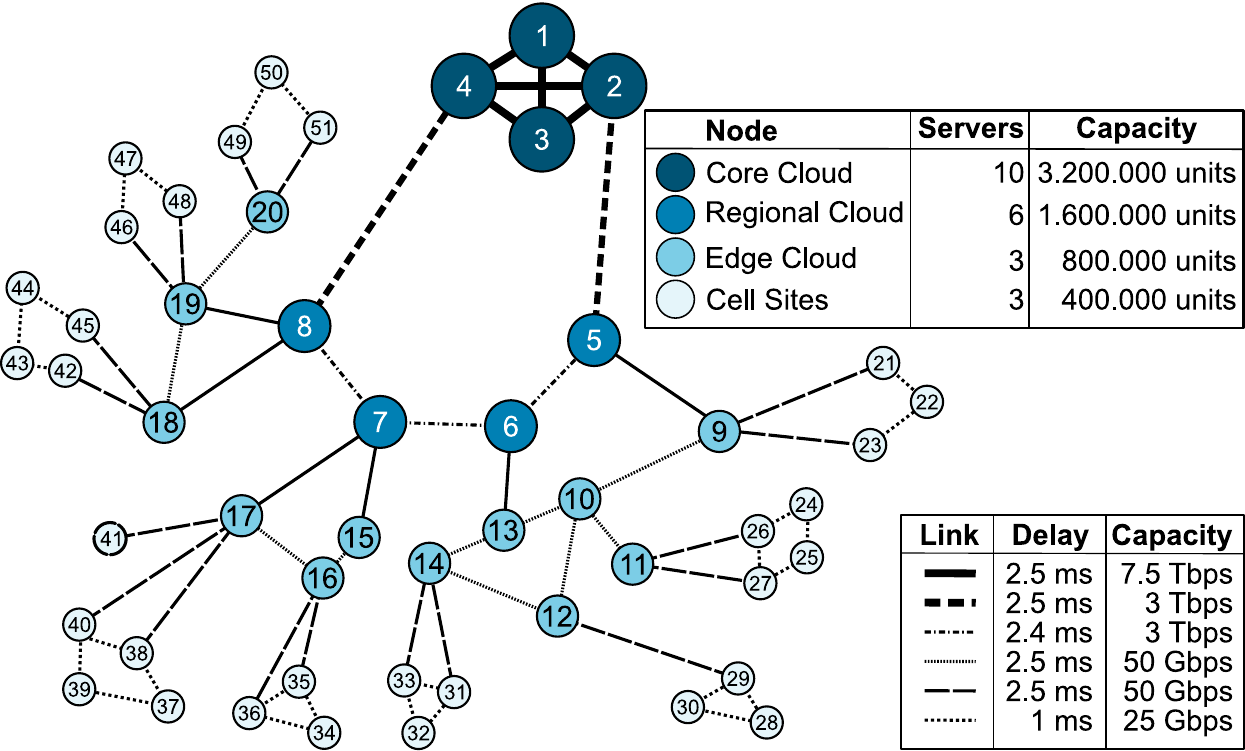}}
\caption{Network topology used in the evaluation.}
\label{fig:topology}
\end{figure}

The evaluation scenario was defined based on real network topologies currently deployed to support RAN~\cite{Morais2022,passion2018}. As depicted in~\figref{fig:topology}, four classes of nodes compose the scenario, and the nodes are classified into tiers. CS (Tier-0) nodes host the RU of the RAN, providing radio interface for UEs of the campus network being the source of each demand. The CSs are physically located near the industrial premises where the campus network should be deployed. Multiple CSs collect mobile application data and feed this data to a smaller number of Edge Clouds (Tier-1), either directly or forwarding the information to other CS sites forming an access ring. The Edge Cloud nodes can host other blocks of the RAN (such as DUs and CUs) or be used as transport nodes. Multiple Edge Clouds are connected to fewer Regional Clouds (Tier-2), which in turn, are connected to a small number of Core Cloud sites (Tier-3). 

All the sites located in a given tier have similar processing capacity, which is given in processing units, and the number of servers. The characteristics of each node regarding processing capacity can be found in \figref{fig:topology} (top table). The considered network is composed in total by 51 nodes connected through 70 bidirectional links characterized by varying capacity and propagation delay according to the tiers that each link connects. The transmission delay and bandwidth of links were derived from deployments of a real metro-area 5G C-RAN~\cite{Liu2022} and the  values assumed per various link types are also specified in~\figref{fig:topology} (bottom table).
 

Each slice deployed in the campus network is composed of a particular set of RAN \acp{NF}. The origin of slices is the \ac{RU} hosted in one of the CSs of the topology depicted in \figref{fig:topology}. Revisting \figref{fig:pni-npn-network}, the network slices in Production line \#2 have a different source (a different RU, possibly hosted in a different CS), therefore the proposed approach considers them as two independent RAN slices. Multiple slices are created for each service type~\cite{itu-m2083} with a subset of slices with the same source node composing a campus network.

Table~\ref{tab:slices} shows the eight slice types considered in the experiments. URLLC\_RO1 and URLLC\_RO2 aim to support communication with high reliability and low latency requirements, such as autonomous robots. URLLC\_AW1 and URLLC\_AW2 are URLLC slices but with improved data rates to support communication for augmented worker devices. Improved experience in mobile broadband services with high data rates is supported by eMBB1 and eMBB2 slice types. Finally, mMTC1 and mMTC2 slice types support machine communication with high device density and relaxed requirements regarding latency and network throughput. Therefore, in our simulated scenario, each mMTC demand is the aggregated demand of a massive number of IoT devices deployed in the industry premises. Although the number of devices is big, the device cycle time and small payload sizes characterize this service type. The values for required bandwidth and total service follow the specifications from~\cite{NGMN-URLLC}, and \cite{5g-acia-mmtc}.

\begin{table}[htbp]
\caption{Slice types and its requirements}
\begin{center}
\begin{tabular}{lrrc}
\toprule
\textbf{Slice type} & \textbf{Bandwidth} & \textbf{Total}& \textbf{RAN} \\ 
\textbf{}           &                    & \textbf{service delay}        & \textbf{isolation} \\ 
\midrule
URLLC\_RO1         & 4~Mbps      &  1~ms    & \checkmark \\ 
URLLC\_RO2         & 25~Mbps     &  1~ms    & \checkmark \\ 
URLLC\_AW1         & 100~Mbps    &  1~ms    & \checkmark \\ 
URLLC\_AW2         & 1~Gbps      &  1~ms    & \checkmark \\ 
eMBB1              & 10~Gbps     &  4~ms    & $-$ \\ 
eMBB2              & 20~Gbps     &  4~ms    & $-$ \\ 
mMTC1              & 1~Mbps      &  15~ms   & $-$ \\ 
mMTC2              & 2~Mbps      &  15~ms   & $-$ \\ 
\bottomrule
\end{tabular}
\label{tab:slices}
\end{center}
\end{table}

RAN nodes have different processing requirements depending on the parts of the RAN protocol stack that its processes. The processing requirements of each \ac{NF} were derived from~\cite{Morais2022}. Based on the parts of the RAN protocol that each \ac{NF} process, a load ratio ($\Gamma_{t(v)}^{pro}$) is assigned according to Table~\ref{tab:vnfprocessing}. 

\begin{table}[htbp]
\caption{Processing characteristics of each NF type }
\begin{center}
\begin{tabular}{llc}
\toprule
\textbf{NF type}&\textbf{RAN Protocols}&\textbf{Load ratio} \\ 
\midrule
CU         & RRC       & 0.9   \\
           & PDCP      &       \\ \midrule
DU         & High RLC  & 1.44  \\
           & Low RLC   &       \\
           & High MAC  &       \\
           & Low MAC   &       \\
           & High PHY  &       \\ \midrule
RU         & Low PHY   & 2.16  \\ 
\bottomrule
\end{tabular}
\label{tab:vnfprocessing}
\end{center}
\end{table}

The set of pre-calculated paths used to route the traffic demands from its origin in the CS to the destination node was generated from all nodes in Tier-0 (\ac{CS}) towards all nodes in Tier-3 (Core Cloud). However, the model allows the last NF of each RAN slice (\ie CU) to be placed in any node along the path.

\section{Results and Verification}
\label{sec:results}

In evaluating the proposed approach, we compare the results of the \ac{MILP} with those from the first-fit algorithm, named in the graphs as LP and HEU, respectively. Both are run considering a public network scenario with the topology depicted in Fig.~\ref{fig:topology}. All the traffic demands of slices have as their source a set of nodes at the CS. As explained previously, we account for a slice's propagation and processing delay from the source of the demand until the respective CU of the RAN transporting the data. The source node for each slice is selected from a uniform distribution over all CS nodes. The type of service that each slice should support is also selected from a uniform distribution, considering the services and characteristics from Table~\ref{tab:slices}. The number of slices supported in the network varies from 5 to 50. The model was solved with the Gurobi Optimizer version 9.5.2. On average, for the highest load (50 slices), each \ac{LP} execution took 13 minutes to complete on an Intel Core i5-10600 CPU at 3.3GHz with 16 GB of RAM with Fedora 37 Workstation. The independent replication method was considered to generate the confidence intervals, with ten executions for each plot point, for a 95\% confidence level. Finally, we assume $\alpha = 0.98$ for all experiments, thus applying significantly more weight to the computation to encourage cheaper computing costs.

\begin{figure}[htbp]
\centerline{\includegraphics[width=0.5\columnwidth]{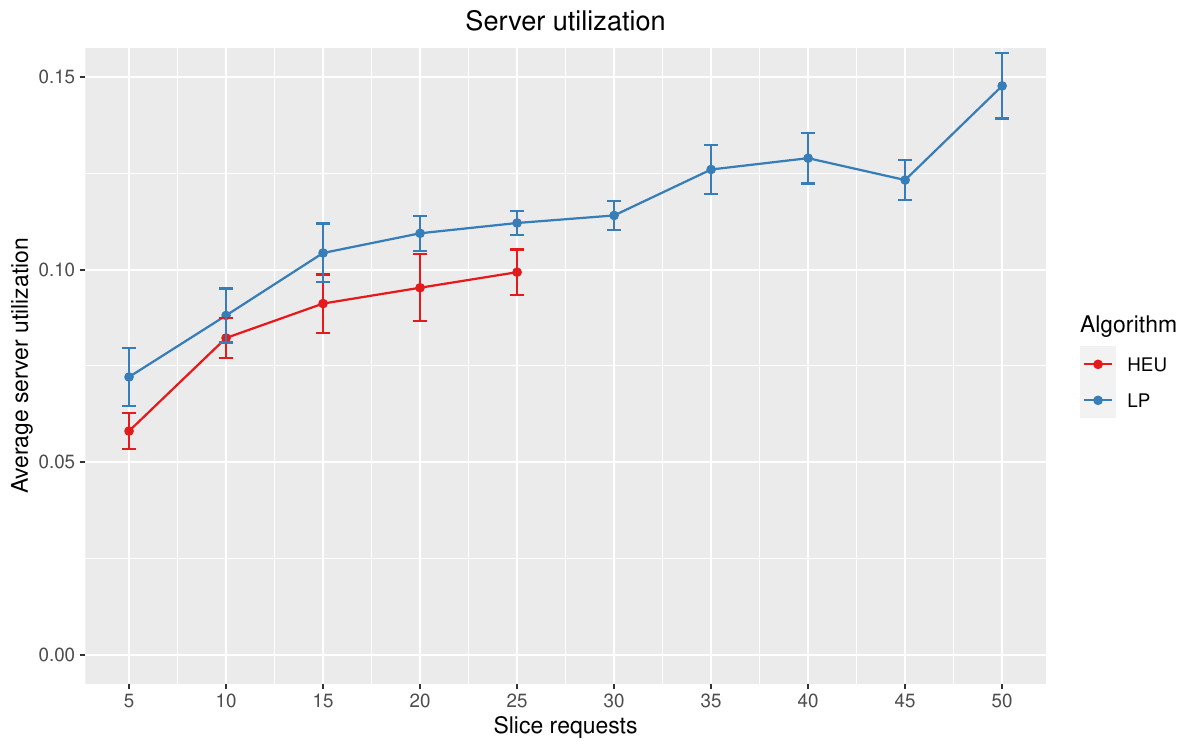}}
\caption{Average server utilization by RAN NFs.}
\label{fig:p1}
\end{figure}

\figref{fig:p1} shows the average server utilization of the RAN blocks (RU, DU and CU). As the first-fit approach could not fulfill the requirements of service delay of slices when placing more than 25 slices in the networks, these results were not considered in the evaluation. On the other hand, the \ac{LP} model was able to distribute the load through the network more efficiently, placing all the slice requests in the network. The model seeks to minimize the computation cost by placing the RAN \acp{NF} in higher tiers of the network (where computational cost is reduced). However, between 30 and 40 slices, due to the topology, we can observe a slight change in the line slope caused by a shortage of computing resources in the regional cloud sites, placing the \acp{NF} of slices with a limited delay budget on the edge cloud and CSs.

\begin{figure}[htbp]
\centerline{\includegraphics[width=0.5\columnwidth]{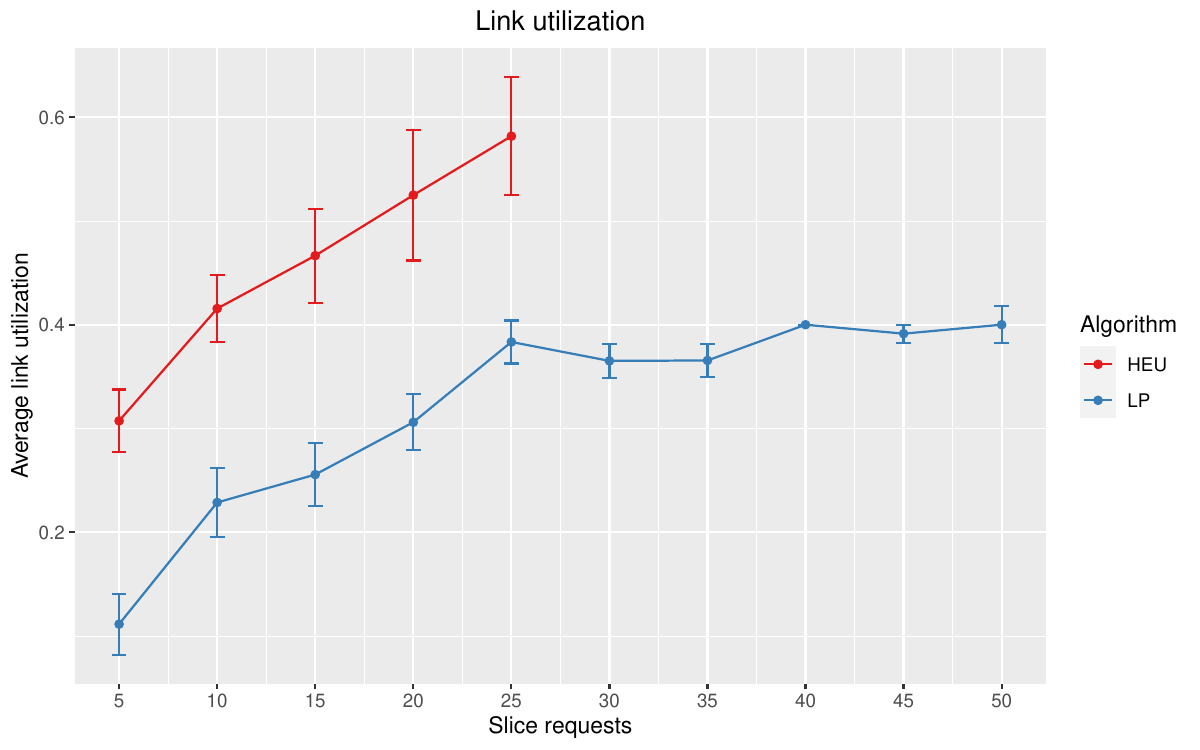}}
\caption{Link utilization.}
\label{fig:p2}
\end{figure}

The average link utilization represented in \figref{fig:p2} corroborates this information showing a reduction in link usage in the region between 30 and 40 slices. Since the \acp{NF} are being placed in lower tiers of the network (close to UEs in the cell site), the average link utilization is reduced. Due to stringent requirements regarding service delay for URLLC and eMBB service types, hindering the use of computing resources at the Core Cloud. Consequently, mobile network operators should increase the computing resource offered at the edge when planning to support many delay-sensitive industrial applications. 

\begin{figure}[htbp]
\centerline{\includegraphics[width=0.5\columnwidth]{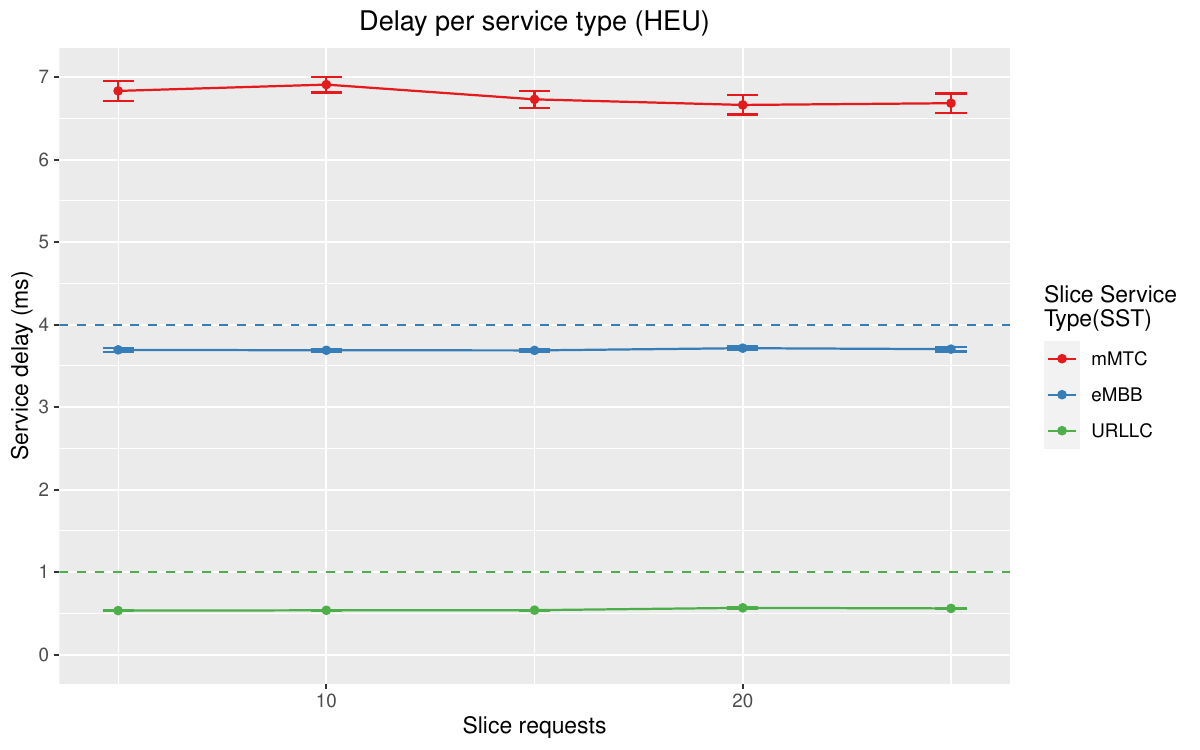}}
\caption{Service delay experienced by the application when placing the NFs using the heuristics algorithm. Horizontal dashed lines indicate the delay budgets.}
\label{fig:p52}
\end{figure}

\begin{figure}[htbp]
\centerline{\includegraphics[width=0.5\columnwidth]{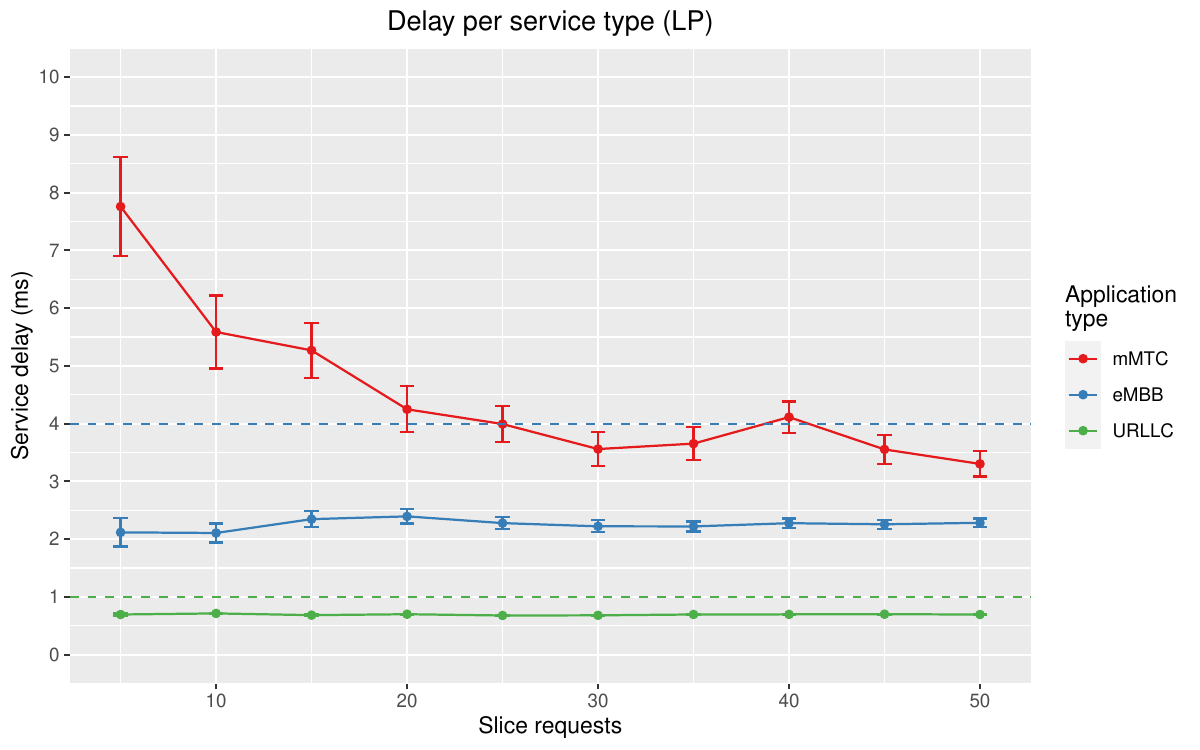}}
\caption{Service delay experienced by the application when placing the NFs using the \ac{LP} model. Horizontal dashed lines indicate the delay budgets.}
\label{fig:p51}
\end{figure}

The delay per service type for the heuristics algorithm plotted in \figref{fig:p52} presents slight variations due to the characteristic of the algorithm that places the RAN \acp{NF} in specific layers according to the slice type. Here it is important to recall that only placements fulfilling the delay budget of the slice are considered in this graph. In other words, placements that do not fulfill the maximum service delay ($D_{s}^{max}$) of the slice are discarded. When comparing the service delay of URLLC slices of the heuristics algorithm with the service delay of URLLC slices of the \ac{LP} model on \figref{fig:p51}, a slight increase in the service delay can be observed. The \ac{LP} model can load balance the processing load over servers in the network, and this behavior is expected. Surprisingly, the service delay of mMTC slices is, on average 3.93, a much smaller value than the maximum service delay allowed for this type of service. Since this type of slice does not require high traffic demands and the processing cost in servers is small, the \ac{LP} model allocated DUs and CUs of this slice type at the regional cloud and the edge cloud without affecting URLLC and eMBB slices due to the balance between the cost of processing and cost of links.

\begin{figure}[htbp]
\centerline{\includegraphics[width=0.5\columnwidth]{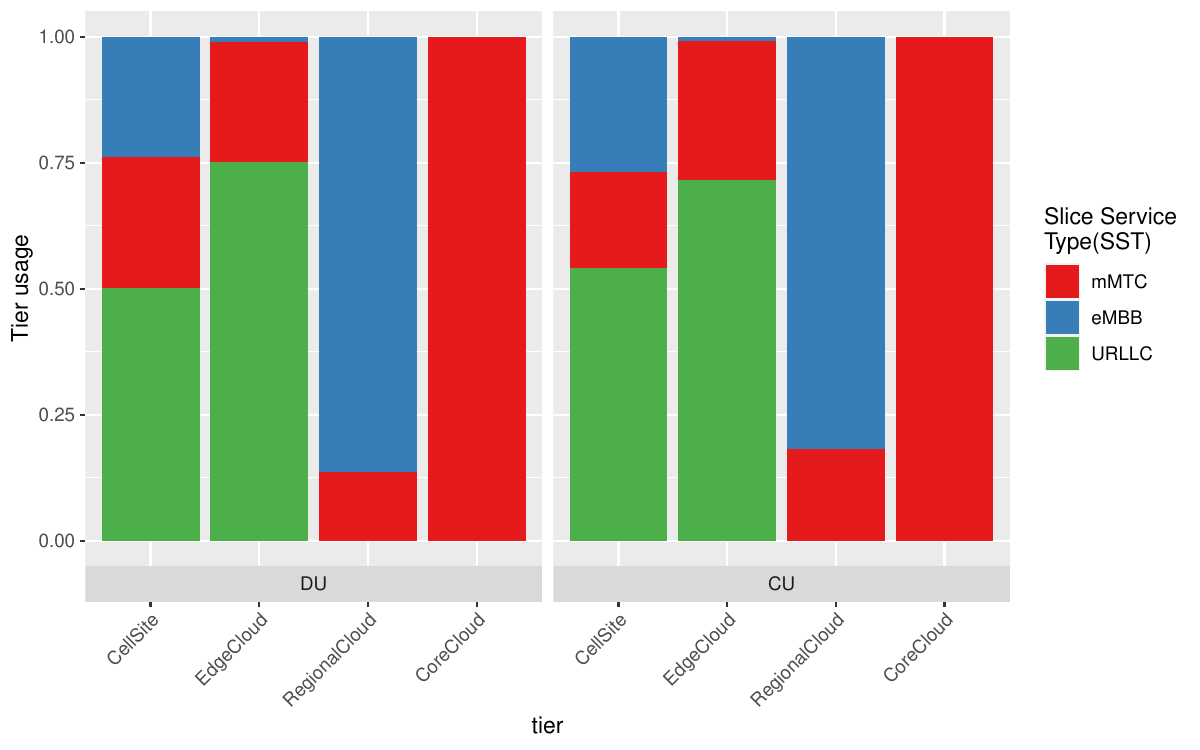}}
\caption{Utilization of servers in each tier by DUs and CUs.}
\label{fig:p6}
\end{figure}

The distribution of DUs and CUs among different network tiers can be observed in \figref{fig:p6}. Due to the service delay requirements of URLLC slices, the corresponding RAN blocks are hosted in the CS and at the Edge Cloud, showing the importance of moving the computing resources to the edge when supporting industrial applications of this type. Moreover, 54.5\% of the \ac{URLLC} slices are completely hosted at the \ac{CS}, such an allocation pattern simplifies the RAN replication in an eventual fault tolerance configuration. 

Finally, as mentioned earlier in Section \ref{sec:rw}, the fact that our model considers the delays introduced by processing and queuing at the server can clearly be seen in \figref{fig:p6}. Slices requiring more network capacity are placed in tiers closer to the user due to higher processing delays, thus reducing the maximum achievable distance between the \acp{UE} and CUs. The regional cloud is mainly occupied by DUs and CUs belonging to \ac{eMBB} slices due to the \ac{E2E} delay and high traffic demand of this type of slice. On the other hand, moving \ac{mMTC} RAN blocks to the edge does not significantly impact the performance of other slices sharing the same physical resources, given its low resource requirements.

\section{Conclusions and Future Work}
\label{sec:conclusion}

We proposed an approach for vertical slicing of the radio access network (RAN) to enable the deployment of multiple and isolated campus networks to accommodate URLLC services. To this end, we modellled RAN function placement as a mixed integer linear programming problem with URLLC-related constraints. We demonstrated that our approach can find optimal solutions in real-world scenarios. 
Furthermore, unlike existing solutions, our model considered the user traffic flow from a known source node on the network's edge to an unknown \textit{a priori} destination node. This flexibility could be explored in industrial campus networks by allowing dynamic placement of \acp{UPF} to serve the \ac{URLLC}. For future work, we plan to study the isolation of slices at server and node levels. Also, replication of \ac{NF} could be explored to improve the reliability of \ac{NG-RAN} when supporting applications requiring \ac{URLLC}. We also plan to extend the model to allow the sharing of CU and DU \acp{NF} to assess the impact on reliablity. 

\section*{Acknowledgment}
The authors acknowledge the financial support by the Federal Ministry of Education and Research of Germany in the programme of ``Souverän. Digital. Vernetzt'' Joint project 6G-RIC, project identification number  16KISK031, and the European Union’s Horizon Europe under Grant Agreement no. 101096342 (HORSE project).

\bibliographystyle{unsrtnat}
\bibliography{references}  

@article{Morais2022,
  author  = {Morais, Fernando Zanferrari and De Almeida, Gabriel Matheus Faria and Pinto, Leizer Lima and Cardoso, Kleber and Contreras, Luis M. and Righi, Rodrigo da Rosa and Both, Cristiano Bonato},
  journal = {IEEE Transactions on Mobile Computing},
  title   = {{PlaceRAN}: Optimal Placement of Virtualized Network Functions in Beyond {5G} Radio Access Networks},
  year    = {2022},
  volume  = {},
  number  = {},
  pages   = {1-1},
  doi     = {10.1109/TMC.2022.3171525}
}

@article{Gomes2020,
  author  = {Gomes, Rafael L. and Bittencourt, Luiz F. and Madeira, Edmundo R. M.},
  journal = {IEEE Communications Magazine},
  title   = {Reliability-Aware Network Slicing in Elastic Demand Scenarios},
  year    = {2020},
  volume  = {58},
  number  = {10},
  pages   = {29-34},
  doi     = {10.1109/MCOM.001.2000753}
}

@article{Yu2020,
  author  = {Yu, Hao and Musumeci, Francesco and Zhang, Jiawei and Tornatore, Massimo and Ji, Yuefeng},
  journal = {Journal of Lightwave Technology},
  title   = {Isolation-Aware {5G RAN} Slice Mapping Over {WDM} Metro-Aggregation Networks},
  year    = {2020},
  volume  = {38},
  number  = {6},
  pages   = {1125-1137},
  doi     = {10.1109/JLT.2020.2973311}
}

@inproceedings{Sarikaya2021,
  author    = {Sarikaya, Egemen and Onur, Ertan},
  booktitle = {2021 17th International Conference on Network and Service Management (CNSM)},
  title     = {Placement of {5G RAN} Slices in Multi-tier {O-RAN} {5G} Networks with Flexible Functional Splits},
  year      = {2021},
  volume    = {},
  number    = {},
  pages     = {274-282},
  doi       = {10.23919/CNSM52442.2021.9615541}
}

@article{Ojaghi2022,
  author  = {Ojaghi, Behnam and Adelantado, Ferran and Verikoukis, Christos},
  journal = {IEEE Transactions on Vehicular Technology},
  title   = {SO-RAN: Dynamic RAN Slicing Via Joint Functional Splitting and MEC Placement},
  year    = {2022},
  volume  = {},
  number  = {},
  pages   = {1-16},
  doi     = {10.1109/TVT.2022.3209069}
}

@article{Coelho2022,
  author  = {da Silva Coelho, Wesley and Benhamiche, Amal and Perrot, Nancy and Secci, Stefano},
  journal = {IEEE Transactions on Network and Service Management},
  title   = {Function Splitting, Isolation, and Placement Trade-Offs in Network Slicing},
  year    = {2022},
  volume  = {19},
  number  = {2},
  pages   = {1920-1936},
  doi     = {10.1109/TNSM.2021.3130915}
}

@techreport{5g-acia-npn,
  title       = {{5G} Non-Public Networks for Industrial Scenarios},
  institution = {{5G} Alliance for Connected Industries and Automation},
  year        = {2019},
  month       = jul,
  address     = {Frankfurt, Germany},
  optauthor   = {{5G ACIA}}
}

@techreport{ORAN-WG6-CAD,
  author     = {{O-RAN Alliance}},
  day        = {31},
  month      = mar,
  year       = {2022},
  intitution = {{O-RAN Alliance}},
  note       = {Version 3.00},
  title      = {Cloud Architecture and Deployment Scenarios for {O-RAN} Virtualized {RAN}}
}

@misc{passion2018,
  author       = {{PASSION Project}},
  howpublished = {\url{https://www.passion-project.eu/wp-content/uploads/2018/10/D2.1_v1.0.pdf}},
  title        = {Definition of Use Cases and Requirements for Network, Systems and Subsystems},
  year         = {2018}
}

@article{Liu2022,
  author  = {Liu, Xiang},
  journal = {Journal of Lightwave Technology},
  title   = {Enabling Optical Network Technologies for {5G} and Beyond},
  year    = {2022},
  volume  = {40},
  number  = {2},
  pages   = {358-367},
  doi     = {10.1109/JLT.2021.3099726}
}

@techreport{itu-m2083,
  optauthor   = {International Telecommunications Union},
  institution = {International Telecommunications Union},
  title       = {{IMT} Vision - Framework and overall objectives of the future development of {IMT} for 2020 and beyond},
  year        = {2015},
  number      = {ITU-R M.2083-0},
  type        = {Rec.}
}

@techreport{NGMN-URLLC,
  title       = {Verticals {URLLC} Use Cases and Requirements},
  institution = {{NGMN Alliance}},
  year        = {2020},
  month       = feb,
  address     = {Frankfurt, Germany},
  optauthor   = {{NGMN Alliance}}
}

@techreport{5g-acia-mmtc,
  title       = {{5G} for Connected Industries and Automation},
  institution = {{5G} Alliance for Connected Industries and Automation},
  year        = {2019},
  month       = feb,
  address     = {Frankfurt, Germany},
  optauthor   = {{5G ACIA}}
}

\end{document}